# Magnon and Phonon Dispersion, Lifetime and Thermal Conductivity of Iron from Spin-Lattice Dynamics Simulations

Xufei Wu[1], Zeyu Liu[1], Tengfei Luo[1,2,*]

1. Aerospace and Mechanical Engineering, University of Notre Dame

2. Center for Sustainable Energy at Notre Dame (ND Energy), University of Notre Dame

* address correspondence to: tluo@nd.edu

**Abstract:**

In recent years, the fundamental physics of spin-thermal (i.e., magnon-phonon) interaction has attracted significant experimental and theoretical interests given its potential paradigm-shifting impacts in areas like spin-thermoelectrics, spin-caloritronics and spintronics. Modelling studies of the transport of magnons and phonons in magnetic crystals are very rare. In this paper, we use spin-lattice dynamics (SLD) simulations to model ferromagnetic crystalline iron, where the spin and lattice systems are coupled through the atomic position-dependent exchange function, and thus the interaction between magnon and phonon is naturally considered. We then present a method combining SLD simulations with spectral energy analysis to calculate the magnon and phonon harmonic (e.g., dispersion, specific heat, group velocity) and anharmonic (e.g., scattering rate) properties, based on which their thermal conductivity values are calculated. This work represents an example of using SLD simulations to understand the transport properties involving coupled magnon and phonon dynamics.

## I. Introduction

In recent years, the fundamental physics of spin-thermal interaction has attracted experimental and theoretical interests[1, 2, 3, 4] given its potential paradigm-shifting impacts in areas related to energy (e.g., spin-thermoelectrics),[5, 6] sensor (e.g., spin-caloritronics) [7, 8, 9] and quantum computing (e.g., spintronics). Magnetic materials are of great interests and have potential applications in nanoscale magnetic devices, due to the recent discovered spin Seebeck effects.[10, 11, 12, 13] The spin-thermal coupling-induced tranductional phenomena are found in a wide range of materials, including metals,[14, 15, 16] semiconductors[11, 17] and insulators.[18, 19, 20, 21] At the microscopic level, the spin-thermal interaction is essentially the interaction between magnons and phonons, both of which are Bosons. The former is the quanta of the spin dynamics and the latter is the quanta of the lattice vibration. Theoretical calculations of phonon transport have been perfected in recent decades,[22, 23, 24] but the transport of magnons, especially under the influence of magnon-phonon interaction, has not receive similar attention.

To explain experimental findings related to spin-thermal coupling-induced phenomena, different phenomenological models have been proposed.[25, 26, 27] However, these models usually only serve the purpose of fitting the data and providing qualitative explanations, lacking predictivity. First-principles calculations of magnon-phonon interactions have long been realized using perturbation methods such as the frozen phonon[28] or frozen magnon[29] approaches. However, these methods are based on the adiabatic treatment of the quasiparticles and have mainly been used to study the effect of magnon on the harmonic properties of phonon (i.e., dispersion) and vice versa. To our knowledge, the scatterings and the resultant scattering rates due to magnon-phonon interaction have not been studied using such perturbation methods.

Spin-lattice dynamics (SLD) is a method that combines molecular dynamics (MD) and spin dynamics by integrating the respective equations of motion during the same time progression. The coupling effect between the spin and lattice is realized by properly formulating the coupling terms in the Hamiltonian, usually through the lattice coordinate-dependent exchange function.[30] Although spins go hand-in-hand with electrons, first-principles calculations show that in reality, even in 'itinerant' ferromagnets, the spin is well-localized to the atomic sites,[31, 32, 33] suggesting the validity of the atomic spin concept, which is the basis of SLD. SLD has been used to study magnetic materials,[32] but most of the works focus on the magnetic properties only (e.g., spin dynamic structure[34, 35] and magnetic susceptibilities[36]).

In this paper, we use equilibrium SLD simulations to model ferromagnetic crystalline body-centered cubic (BCC) iron, where the spin and lattice systems are coupled through the atomic position-dependent exchange function. We then present a method using spectral energy density (SED) analysis to calculate the magnon and phonon harmonic (e.g., dispersion, specific heat, group velocity) and anharmonic (e.g., scattering rate) properties, based on which their thermal conductivity values are further calculated.

**II. Spin-Lattice Model and Simulation Details**

In this work, we adopt a SLD model built by Ma and coworkers.[35, 37, 38, 39, 40, 41] In this model, the coupled spin-lattice system is described by the Heisenberg Hamiltonian with further inclusion of the longitudinal fluctuation of spins described by the Heisenberg-Landau expansion. The complete Hamiltonian used in our study is written as:[37, 42, 43, 44]

$$H_{sp-la} = \sum_i \frac{P_i^2}{2m_i} + U(\{R_i\}) - \frac{1}{2}\sum_{\substack{i,j \\ i \neq j}} J_{ij}(|R_i - R_j|) S_i \cdot S_j + \sum_i (AS_i^2 + BS_i^4 + CS_i^6) \quad (1)$$

Here $P_i$ and $R_i$ are respectively the momentum and position of atom $i$, $U(\{R_i\})$ is the interatomic potential of the lattice, $J_{ij}$ is the magnetic exchange function, $S_i$ is the spin vector of atom $i$. Since $J_{ij}$ is a function of the atomic coordinate, phonon and magnon coupling is inherent in this third term.[45] $AS_i^2 + BS_i^4 + CS_i^6$ are the Landau expansion terms,[46, 47, 48] describing how energy varies as a function of magnitude of spin. It is noted that due to the time reversal symmetry, the order of all spins has to be even.[49] The Landau expansion terms are usually not considered in SLD simulations due to the higher order nature.[32, 35, 37, 43] However, it may be found necessary for some systems to correctly reproduce certain magnetic properties, such as the distribution of the magnitudes of magnetic moments.[38]

The equations of motion for the spin and lattice degrees of freedom can be derived from the above Hamiltonian:

$$\frac{dR_i}{dt} = \frac{P_i}{m_i} \ ; \ \frac{dP_i}{dt} = -\frac{\partial U}{\partial R_i} + \sum_j \frac{\partial J_{ij}}{\partial R_i} S_i \cdot S_j \ ; \ \frac{dS_i}{dt} = \frac{1}{\hbar} S_i \times H_i \quad (2)$$

where $H_i = -\frac{dH_{sp-la}}{dS_i} = \sum_j J_{ij} S_i - 2AS_i - 4BS_i^3 + 6CS_i^5$ is the effective exchange field on spin $S_i$. It is noted that the Hamiltonian in Eq. 1 describes a coupled spin-lattice system and the electron degrees of freedom are not included. Due to the significant difference in time scale between electron and spin-lattice systems, it is not practical to directly include electron movement in SLD simulations. An effective solution is to implement Langevin dynamics to the second and the third equations in Eq. 2 to simulate the effect from the electron subsystem (Eq. 3). This scheme has been successfully implemented to study the magnetic properties of iron using SLD.[37]

$$\frac{dP_i}{dt} = -\frac{\partial U}{\partial R_i} + \sum_j \frac{\partial J_{ij}}{\partial R_i} S_i \cdot S_j - \frac{\gamma_l}{m_i} P_i + f_i \; ; \; \frac{dS_i}{dt} = \frac{1}{\hbar} S_i \times H_i + \gamma_s' H_i + \xi_i \qquad (3)$$

here $\gamma_l$ and $\gamma_s'$ are the electron-spin and electron-lattice damping parameters, respectively, assuming the fluctuation and dissipation are conducted by electrons. $f_i$ and $\xi_i$ are Gaussian random variables and its Cartesian components have means of 0 and variance determined according to the electron temperature and damping parameters:

$$\langle \xi_{i\alpha}(t) \rangle = \langle f_{i\alpha}(t) \rangle = 0 \; ; \; \langle \xi_{i\alpha}(t) \cdot \xi_{i\alpha}(t + \Delta t) \rangle = 2\gamma_s' k_B T_e \delta_{ij} \delta_{\alpha\beta} \delta(\Delta t) \; ;$$

$$\langle f_{i\alpha}(t) \cdot f_{i\alpha}(t + \Delta t) \rangle = 2\gamma_l k_B T_e \delta_{ij} \delta_{\alpha\beta} \delta(\Delta t) \qquad (4)$$

where $\alpha$ and $\beta$ denote $x, y,$ and $z$.

In this simulation, we follow Ref. [37] to model the exchange function of BCC iron as $J(r) = J_o(1 - r/r_c)^3 \Theta(r_c - r)$, where $J_o$ =749.588 meV, $r_c$=3.75 Å, and $\Theta$ is the Heaviside step function. The electron-spin and electron-lattice damping parameters, $\gamma_l/m$=0.6 ps$^{-1}$ and $\gamma_s'$=5.88×10$^{13}$ eV$^{-1}$s$^{-1}$, are also adopted from Ref. [37], which was obtained by fitting to the ultrafast demagnetization laser experiments in iron films. The coefficients of the Heisenberg-Landau terms are $A$=-0.440987 eV, $B$=0.150546 eV, and $C$=0.0506794 eV, which were obtained from fitting to first-principles calculations.[38] The SLD simulation is performed on a BCC iron crystal with a size of 45.86$^3$ Å$^3$. We use the Suzuki-Trotter decomposition (STD) integration algorithm to integrate the equations of motion as detailed in Ref. [39]. The interatomic potential used is the embedded atom method (EAM) potential.[50] The simulations are carried out using an in-house code. The simulation temperature is set to be 300 K, and periodic boundary conditions are applied to all directions. The time step is set to 1 fs.

**III. Dispersion Relations**

We first calculate the dispersion relations of phonons and magnons by solving their respective dynamical equations.[28, 51] For phonons, the harmonic interatomic force constants needed to construct the dynamical matrices are obtained from a set of force-displacement data obtained by systematically displacing atoms by a small amount from their equilibrium positions and calculating the resultant forces on atoms. The calculated phonon dispersion is shown in **Fig. 1**a as solid lines. This dispersion agrees well with experimental data in Ref. [52]. For magnon dispersion, we solve the dynamical equation for the spin system:

$$\hbar\omega = \sum_{i \neq o} J(|\mathbf{R}_o - \mathbf{R}_i|)\left(1 - e^{i\mathbf{k}\cdot(\mathbf{R}_o - \mathbf{R}_i)}\right) \quad (5)$$

where $\mathbf{k}$ is the wavevector. This equation is solved analytically, e.g., along the reciprocal [001] direction, the dispersion relation is:

$$\hbar\omega = 8J_1\left(1 - \cos\left(\frac{ka}{2}\right)\right) + 2J_2(1 - \cos(ka)) \quad (6)$$

where $J_1$ and $J_2$ are the magnetic exchange functions with the first and second nearest neighbors, respectively, and $a$ is the lattice constant. In the long wavelength limit, $ka \ll 1$, we have $\omega = (J_1 + J_2)\frac{a^2}{\hbar}k^2$, and thus magnons have a quadratic dispersion near the zone center, which is different from phonons which have linear dispersions for the acoustic branches in the long wavelength limit. The calculated magnon dispersion is shown in Fig. 1b. It is noted that both the quadratic dispersion and the frequency range of the magnon agree favorably with previous first-principles calculations[53, 54] and data from neuron scattering experiments,[55] although a slight over estimation of the frequency is seen. The maximum frequency of magnons is more than 10 times higher than that of phonons. We have also calculated the density of states (DOS) of both phonons and magnons as plotted next to the dispersions. It is seen that phonons have larger DOS than magnons since phonons have three polarizations while magnons only have one and magnon modes are spread over a much larger frequency range (0-105 THz) compared to phonons (0-9.7 THz).

**IV. Specific Heat Capacity and Group Velocity**

Since both phonon and magnon are Bosons, we calculate the specific heat capacity based on the Bose-Einstein distribution. The calculated modal specific heat at 300 K is shown in **Fig. 2**a (blue line). As can be seen, specific heat drops very fast as frequency increases, and modes beyond 60 THz does not carry much energy at 300 K. Since the maximum phonon frequency in our study is ~ 9.7 THz, specific heat of phonon modes is always above 1000 kJ/m$^3$K. This is not the case for magnons since their frequencies range up to ~105 THz, and magnons with frequencies lower than 20 THz contribute more than half of the total heat capacity at 300 K (inset in Fig. 2a). Due to the large difference in DOS (Fig. 1) and that high frequency modes do not contribute much to heat capacity, the cumulative specific heat (Fig. 2a) of magnon (~40 kJ/m$^3$K) is about two orders of magnitude smaller than that of phonon (~3300 kJ/m$^3$K).

The group velocities of magnons and phonons can be calculated by taking the gradients of the dispersion. In Fig. 2b, we plot the group velocities of phonons and magnons as a function of frequency. As both phonon and magnon share the save crystal structure and have the same wave vectors, the large difference between phonon and magnon group velocities is directly from the difference in frequencies. It is understandable that the group velocities of magnons are about 10-20 times greater than those of phonons due to the one order of magnitude difference in their peak frequencies (Fig. 1).

**IV. Phonon and Magnon Lifetimes from Spectral Density Analysis**

Lifetime is another important parameter that determines the transport of quasi-particles. Phonon lifetimes can be computed from the SED analysis using MD data,[56] which has been used to study the lifetime of phonons for different materials.[23, 24, 56, 57] The phonon SED is calculated by taking the Fourier transform in space of the power spectral density of atoms:[56]

$$\psi(\boldsymbol{k}, \omega) \propto \frac{1}{2}\sum_{\alpha\in\{x,y,z\}} m \left| \int_0^{\tau_0} \sum_{n_{x,y,z}} \dot{\boldsymbol{u}}_\alpha(n_{x,y,z}, b; t) \times exp[i\boldsymbol{k}\cdot \boldsymbol{R}(n_{x,y,z}) - i\omega t] dt \right|^2 \quad (7)$$

where $n_{x,y,z}$ represents different unit cells, and $\boldsymbol{R}(n_{x,y,z})$ represents the vector of displacement between unit cell $n_{x,y,z}$ and the basis unit cell. $\psi(\boldsymbol{k}, \omega)$ can be fitted by a linear superposition of Lorentzian functions, where the peaks are phonon frequencies and the inverse of the linewidth is the phonon lifetime. **Figure 3**a shows sample SED peaks of phonons and an example Lorentzian fit. The phonon frequencies obtained from the SED analysis are shown in Fig. 1a as blue stars. As can be seen, the frequencies agree very well with the dispersion calculated from the dynamical equations, especially at low frequencies. High frequency phonon modes are compressed, which may be attributed to the phonon-magnon interaction in the SLD simulations. However, since these phonons have smaller group velocities and heat capacity compared to those of low frequency modes, the small discrepancy in the high frequency region will not impact the overall transport of phonons.

For magnons, a similar route is taken. We can decompose a magnon mode into the steady (ST) and transient (T) parts:

$$S(t|\mathbf{k}, v) = S_{ST}(t|\mathbf{k}, v) + S_T(t|\mathbf{k}, v) \tag{8}$$

where

$$S_{ST}(t|\mathbf{k}, v) = C_1(\mathbf{k}, v)e^{i\omega(\mathbf{k},v)t} + C_2(\mathbf{k}, v)e^{-i\omega(\mathbf{k},v)t} \tag{9}$$

$$S_T(t|\mathbf{k}, v) = e^{-\Gamma(\mathbf{k},v)t}\big(C_3(\mathbf{k}, v)e^{i\omega(\mathbf{k},v)t} + C_4(\mathbf{k}, v)e^{-i\omega(\mathbf{k},v)t}\big) \tag{10}$$

Here $v$ is polarization (note: there is only one polarization for BCC iron magnon), and $C_i$ are constants, and $\Gamma$ is the half-width at half-maximum of the SED peak corresponding to a magnon mode. Similar to phonons, we formulate the spectral energy for magnons:

$$\phi(\mathbf{k}, \omega) \propto \sum_{\alpha \in \{x,y,z\}} \sum_b \hbar \left| \int_0^{\tau_0} \sum_{n_{x,y,z}} \frac{dS_\alpha(n_{x,y,z},b;t)}{dt} \times exp[i\mathbf{k} \cdot \mathbf{R}(n_{x,y,z}) - i\omega t] dt \right| \tag{11}$$

It can be shown that $\phi(\mathbf{k}, \omega)$ is the spin energy of a magnon with wavevector $\mathbf{k}$ and frequency $\omega$. Similar to the proof of phonon SED,[56] we can show that $|\phi|^2(\mathbf{k}, \omega)$ can be expressed as the superposition of Lorentzian functions:

$$|\phi|^2(\mathbf{k}, \omega) = \sum_v \frac{I_v}{[(\omega - \omega_v)/\Gamma(\mathbf{k},v)]^2 + 1} \tag{12}$$

where $\omega_v$ is the frequency at the peak center, $I_v$ is the peak magnitude. The magnon lifetime can be calculated as $\tau = 1/(2\Gamma)$. Figure 3b shows sample magnon SED peaks. The frequencies extracted from the peaks are plotted in Fig. 1b as red circles. It is seen that the SED calculated frequencies agree almost perfectly with those calculated analytically from the dynamical equation. There is slight over-estimation of the SED frequencies compared to the ones from the dynamical equation at high frequencies. We have found that this is related to the high order Landau terms, $AS_i^2 + BS_i^4 + CS_i^6$, because when we performed SED analysis of SLD simulations without such terms, the dispersion curve at the high frequency region showed lower frequencies than those of the analytical dispersion. Again, such small discrepancy will not influence the description of the overall energy transport by magnons since the high frequency modes have small velocities and are not even excited at 300 K (Fig. 2a). As a result, the SED analysis of a SLD simulation trajectory provides an alternative way to calculate the magnon dispersion relation, which is commonly obtained from the dynamical equations at the ground state or dynamical structure factor measured in neutron scattering experiments.[33, 58, 59]

From the linewidth, we also calculate the lifetimes of magnons. As shown in Fig. 3c, our calculated lifetimes are on the right order of magnitude compared to the only available first-principles calculations,[60] although our currently used SLD Hamiltonian was not initially designed to predict this quantity. In Fig. 3c, lifetimes of magnons and phonons have the same decreasing trends as frequency increases. It is found that the lifetimes of magnons follow a relation of $\omega^{-1}$. In the meantime, the phonon lifetimes are more spread in trend and seem to possess a scaling between $\omega^{-1}$ and $\omega^{-2}$. Such a trend of phonon lifetimes is similar to that found in other crystals such as half-Heusler.[23]

**V. Thermal Energy Transport by Phonon and Magnon**

With the specific heat, group velocity and lifetime obtained from above, we can further calculate the thermal conductivity, $\kappa$, of phonons and magnons according to the kinetic theory, $\kappa = \frac{1}{3N_q} \sum_q c_{qs} v_{qs}^2 \tau_{qs}$, where $N_q$ is the total number of discrete grid points in the first Brillouin zone. The calculated thermal conductivity of phonon and magnon turns out to be 8.3 and 15.2 W/mK, respectively. The experimental thermal conductivity of iron is ~77.3 W/mK,[61] and the contribution from electron is calculated to be 69.4 W/mK according to Wiedemann-Franz law, leaving the contributions from phonons and magnons to be ~7.9 W/mK. Conventionally, this part has been attributed to phonons,[61] and this would agree very well with our value of 8.3 W/mK. Although there is no experimental data to compare to according to our knowledge, it is still somewhat surprising to see that magnon has a large thermal conductivity of 15.2 W/mK. This may be partially attributed to the accuracy of the SLD model which is still empirical and involving parameters fitted to experimental data. It is seen from Fig. 1b that the SLD model might have over predicted the magnon group velocity compared to experimental dispersion. A rough estimation indicates that the over-prediction is about 50%. If we take such a factor into account, the thermal conductivity will reduce from 15.2 W/mK to ~6.8 W/mK. Our calculated lifetimes (Fig. 3c) are also ~ 2 times larger than first-principles calculations.[60] However, even further considering this factor in addition to the group velocity factor, the thermal conductivity will be ~3.4 W/mK, which is still on the same order of magnitude as that of phonon. Adding this magnon thermal conductivity to those from phonon and electron, the total value would be ~81.1 W/mK, which is about 5% higher than reference value of 77.3 W/mK. We deem such agreement to be reasonable, especially considering defect scatterings of phonons and magnons in real samples. The defect scattering of magnons was shown to scale with the 4$^{th}$ power of the wavevector and be proportional to the defect density,[62] but there is no way to know the defect concentrations in the experimental samples. Of course, more accurate models that can precisely predict the dispersions and lifetimes of both phonons and magnons are desirable for future studies.

## VI. Conclusions

In summary, we use SLD simulations to model ferromagnetic crystalline iron, where the spin and lattice systems are coupled through the atomic position-dependent exchange function. Dispersion relations calculated from dynamical equations agree well with those from SED analysis. The specific heat of the phonon system is about two orders of magnitude greater than that of magnons largely due to the difference in their DOS. The group velocities of magnons are more than 10× larger than those of phonons, but their lifetimes are about one order of magnitude smaller than those of phonons. The calculated thermal conductivity values for phonons and magnons are on the same order of magnitude. This work represents an important step towards the simulation understanding of the transport properties involving coupled magnon and phonon dynamics, and it may inspire new research activities in this field.

**Acknowledgements**

The authors would like to thanks the support from NSF (1433490). The simulations are supported by the Notre Dame Center for Research Computing, and NSF through XSEDE computing resources provided by SDSC Comet and Comet and TACC Stampede under grant number TG-CTS100078. T.L. also wants to thank the endowed professorship from the Dorini Family.

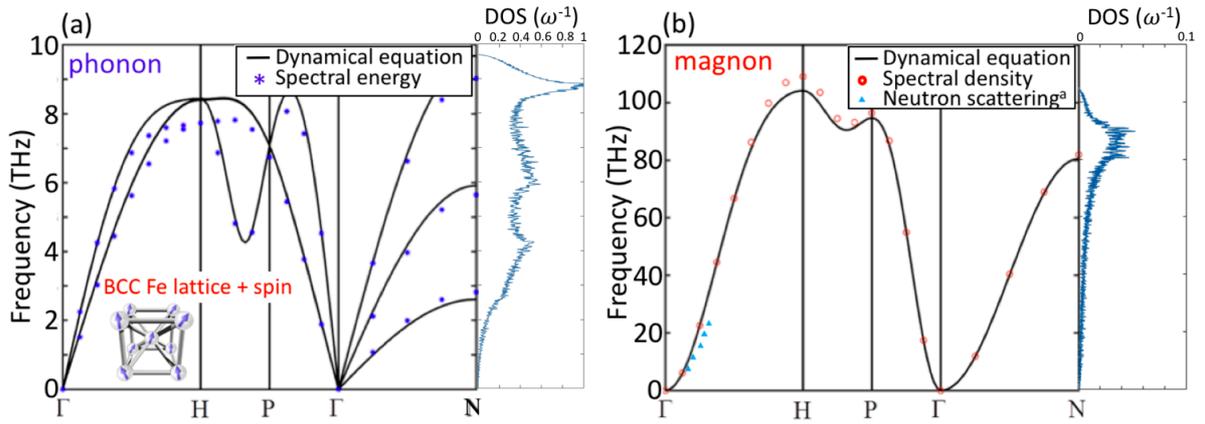

**Figure 1**. (a) Phonon dispersion and density of state (DOS) and (b) magnon dispersion relations and DOS of BCC iron. Black solid lines are dispersions calculated from dynamical equations, and blue stars and red circles are phonon and magnon frequencies fitted from spectral energy density and spectral density methods, respectively. Blue triangles in panel (b) are neutron scattering data from Ref. [55]. The inset in panel (a) is a schematic of the BCC iron structure with atomic spins superimposed on the lattice sites. The DOS are divided by the number of grid sampling points in the Brillouin zone, which are chosen to be the same for both magnon and phonon calculations.

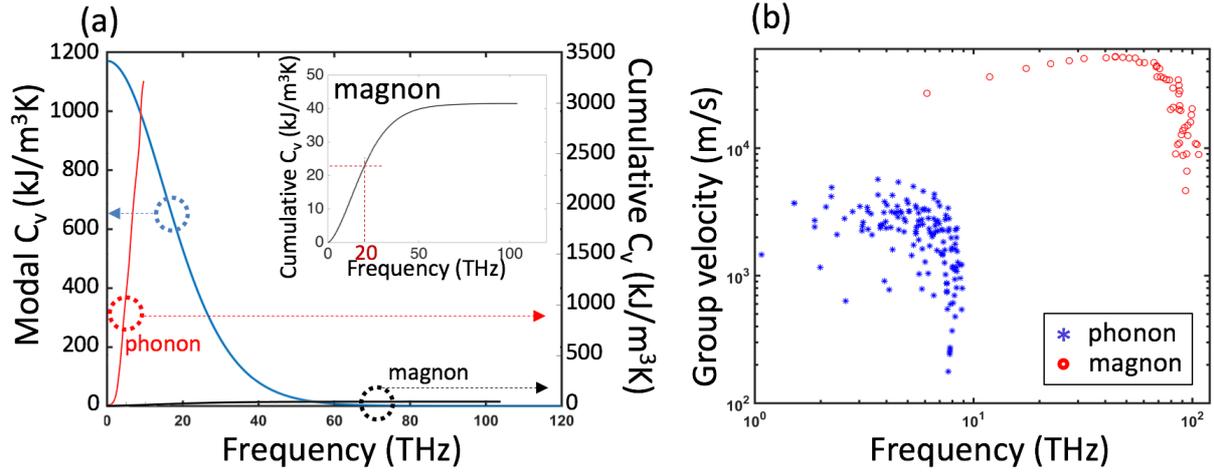

**Figure 2**. (a) Specific heat as a function of frequency (blue, left axis), and cumulative specific heat of phonon (red, right axis) and magnon (black, right axis). Inset: blow-up view of the cumulative specific heat of magnon. (b) Group velocities of magnon and phonon modes calculated from the dispersion relation.

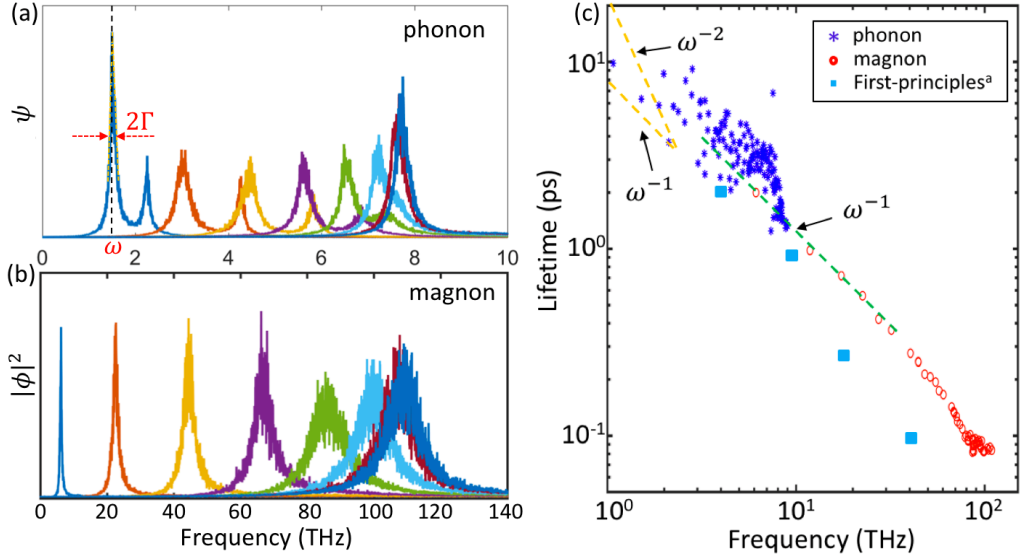

**Figure 3**. Spectral energy of (a) phonons and (b) magnons. In panel (a), an example Lorentzian fit with the linewidth ($\Gamma$) and frequency ($\omega$) is shown. (c) Phonon and magnon lifetimes calculated from the linewidth of the Lorentzian fits. Magnon lifetimes scale well with $\omega^{-1}$ while those of phonons scale with $\omega^{-1} \sim \omega^{-2}$ in the low frequency limit. First-principles results are from Ref. [60].